\def\build#1_#2^#3{\mathrel{
\mathop{\kern 0pt#1}\limits_{#2}^{#3}}}

\newcommand{\be}{\begin{equation}}
\newcommand{\ee}{\end{equation}}
\newcommand{\bea}{\begin{eqnarray}}
\newcommand{\eea}{\end{eqnarray}}

\documentstyle[prl,aps]{revtex}
\begin{document}
\twocolumn[\hsize\textwidth\columnwidth\hsize\csname@twocolumnfalse%
\endcsname
\rightline{IHES/P/02/27 , Bicocca-FT-02-7, CERN-TH/2002-092}
\vspace{3mm}

\draft
%
%
\title{Runaway dilaton and equivalence principle violations}
\author{Thibault Damour}
\address{
Institut des Hautes Etudes Scientifiques, 35 route de Chartres, 
F-91440 Bures-sur-Yvette, France
}
\author{Federico Piazza}
\address{
Dipartimento di Fisica, Universita di Milano Bicocca, 
Piazza delle Scienze 3, I-20126 Milan, Italy
}
\author{Gabriele Veneziano}
\address{
Theory Division, CERN, CH-1211 Geneva 23, Switzerland, and \\
Laboratoire de Physique Th\'eorique, Universit\'e Paris-Sud, 91405 Orsay, 
France
}
\maketitle

\begin{abstract}
{
In a recently proposed scenario, where the dilaton decouples while  cosmologically attracted 
towards infinite bare string coupling, its
residual  interactions
can be related to the amplitude of density fluctuations generated during 
inflation,  and are large enough to  be detectable through a modest improvement 
on present tests  of free-fall universality. Provided it has 
 significant couplings to either dark matter or dark energy, a runaway dilaton can also
induce time-variations of the natural ``constants"  within the reach of near-future
 experiments.
}
\end{abstract}
]

A striking prediction of all string theory models is the existence of a scalar partner of the spin 2 
graviton: the dilaton $\phi$, whose vacuum expectation value (VEV) determines the 
string coupling constant $g_s = e^{\phi / 2}$ \cite{Witten}. At tree level,
 the dilaton is massless 
and has gravitational-strength couplings to matter which violate the equivalence 
principle \cite{TV88}. This is in violent conflict with present experimental tests of 
general relativity. It is generally assumed that this conflict is avoided because, 
after supersymmetry breaking, the dilaton might acquire a (large enough) mass (say 
$m_{\phi} \gtrsim 10^{-3} \, {\rm eV}$ so that observable deviations from Einstein's 
gravity are quenched on distances larger than a fraction of a millimeter). However, 
Ref.~\cite{DP94} (see also \cite{DN93}) has proposed a mechanism which can naturally 
reconcile a {\it massless} dilaton with existing experimental data. The basic idea of 
Ref.~\cite{DP94} was to exploit the string-loop modifications of the (four dimensional) 
effective low-energy action (we use the signature $-+++$)
\begin{eqnarray}
\label{eq1.1}
S & = & \int d^4 x \sqrt{\widetilde{g}} \, \biggl( \frac{B_g (\phi)}{\alpha'} \, \widetilde R + 
\frac{B_{\phi} (\phi)}{\alpha'} \, \lbrack 2 \widetilde{\build\Box_{}^{}} \phi -
(\widetilde{\nabla} \phi)^2 \rbrack 
 \nonumber \\
& & - \frac{1}{4} \, B_F (\phi) \, \widetilde{F}^2 - \dots \biggl) \, ,
\end{eqnarray}
i.e. the $\phi$-dependence of the various coefficients $B_i (\phi)$, $i = g , \phi , F, 
\ldots$ , given in the weak-coupling region ($e^{\phi} \to 0$) by series of the form
$B_i (\phi) = e^{-\phi} + c_0^{(i)} + c_1^{(i)} \, e^{\phi} + c_2^{(i)} \, e^{2\phi} + 
\cdots \, $,
coming from genus expansion of string theory. It was shown in \cite{DP94} that, if there exists a special 
value $\phi_m$ of $\phi$ which extremizes all the (relevant) coupling functions 
$B_i^{-1} (\phi)$, the cosmological evolution of the graviton-dilaton-matter system 
naturally drives $\phi$ towards $\phi_m$ (which is a fixed point of the 
Einstein-dilaton-matter system). This provides a mechanism for fixing a 
massless dilaton at a value where it decouples from matter (``Least Coupling 
Principle''). In this letter, we consider the case (recently suggested in \cite{V01})
where the coupling functions, at least in the visible sector,
 have a smooth {\it finite} limit for infinite 
bare string coupling  $g_s \to \infty$. In this case, quite
generically, we expect
\be
\label{eq1.3}
B_i (\phi) = C_i + {\cal O} (e^{-\phi}) \, .
\ee
Under this assumption, the coupling functions are all extremized at infinity, i.e. 
a fixed point of the cosmological evolution is $\phi_m = + \infty$. [See \cite{GPV}
for an exploration of the late-time cosmology of models satisfying (\ref{eq1.3}).]
We found that the ``decoupling'' of such a ``run-away'' dilaton   
has remarkable features: (i) the residual dilaton couplings at the
present epoch can be related to the amplitude of density fluctuations generated
during inflation, and (ii) these residual couplings, while being naturally
compatible with present experimental data, are predicted to be large enough to
be detectable by a modest improvement in the precision of equivalence principle
tests (non universality of the free fall, and, possibly, variation of ``constants'').
This result contrasts with the case of attraction towards a finite
value $\phi_m$ which leads to extremely small residual couplings \cite{DV96}.

We assume some primordial inflationary stage driven by the potential energy of an
inflaton field $\chi$. Working with the Einstein frame metric
$g_{\mu \nu} = C_g^{-1} \, B_g (\phi) \, \widetilde{g}_{\mu \nu}$, and with the modified
dilaton field 
$\varphi = \int d \phi [ (3/4) ( B'_g/B_g )^2 + 
B'_{\phi}/B_g + (1/2) \, B_{\phi}/B_{g} ]^{1/2}$,
 we consider an effective action of the form
\begin{eqnarray}
\label{eq2.4}
S & = & \int d^4 x \sqrt{g} \biggl[ \frac{\widetilde{m}_P^2}{4} \, R - 
\frac{\widetilde{m}_P^2}{2} \, (\nabla \varphi)^2
 \nonumber \\
& & - \frac{\widetilde{m}_P^2}{2} \, F 
(\varphi) (\nabla \chi)^2 - \widetilde{m}_P^4 \, V (\chi , \varphi) \biggl] \, ,
\end{eqnarray}
where $\widetilde{m}_P^2 = 1/(4\pi G) = 4 C_g/ \alpha'$, and where the dilaton
dependence of the Einstein-frame action is related to its (generic) string-frame
dependence (\ref{eq1.1}) by 
$F(\varphi) = B_{\chi} (\phi) / B_g (\phi) \, , \ V(\chi , \varphi) = C_g^{2} \, 
\widetilde{m}_P^{-4} \, B_g^{-2} (\phi) \, \widetilde V (\widetilde{\chi} , \phi)$.

Under our basic assumption (\ref{eq1.3}), 
$d\varphi / d \phi$ tends, in the strong-coupling limit $\phi \to +\infty$,
to the constant $1/c$, with $c \equiv (2 C_g / C_{\phi})^{1/2}$, so that the
asymptotic behaviour of the bare string coupling is 
\be
\label{eq2.6}
g_s^2 = e^{\phi} \simeq e^{c\varphi} \, .
\ee
Let us consider the  case where $F(\varphi) = 1$ and 
$V (\chi , \varphi) = \lambda (\varphi) \, \chi^n/n$ 
 with a dilaton-dependent inflaton coupling constant $\lambda (\varphi)$  of the form
\be
\label{eq2.11}
\lambda (\varphi) = \lambda_{\infty} (1 + b_{\lambda} \, e^{-c\varphi}) \, ,
\ee
where we assume that $b_{\lambda} > 0$, 
i.e that $\lambda (\varphi)$ reaches a {\it minimum} at strong-coupling, $\varphi 
\rightarrow + \infty$.
It is shown in \cite{DPVlong} that this simple case is 
representative of rather general cases of $\varphi$-dependent inflationary potentials 
$V(\chi , \varphi)$

During  inflation ($ds^2 = -dt^2 + a^2 (t) \, \delta_{ij} \, dx^i \, 
dx^j$), it is easily seen that, while $\chi$ slowly rolls down towards
$\chi \sim 1$, the dilaton $\varphi$ is monotically driven towards large values.
The solution of the (classical) slow-roll 
evolution equations leads to  
\be \label{eq2.16} 
e^{c \varphi} + \frac{b_{\lambda} \, c^2}{2n} \, \chi^2 = {\rm const}. = e^{c 
\varphi_{\rm in}} + \frac{b_{\lambda} \, c^2}{2n} \, \chi_{\rm in}^2 \, .
\ee
Using the result (\ref{eq2.16}), we estimate the value $\varphi_{\rm end}$
of $\varphi$ at the end
of inflation by inserting for the initial value $\chi_{\rm in}$ of the inflaton
the value corresponding to the end of self-regenerating inflation \cite{L90}.
We remark that the latter value can be related to the amplitude 
${\delta_H} \sim 5 \times 10^{-5}$
of density fluctuations, on the scale corresponding to our present horizon,
generated by inflation, through
$\chi_{\rm in} \simeq \, 5 \sqrt n \, 
({\delta_H} )^{ - 2/(n+2)}$. Finally, 
assuming $e^{c \varphi_{\rm in}} \ll e^{c \varphi_{\rm end}}$,
we get the estimate:
 \be
\label{eq2.23'}
e^{c \varphi_{\rm end}}  \sim
 12.5  c^2 \, b_{\lambda} \, 
\left(\delta_H\right)^{ -\frac{4}{n+2}} \, .
\ee
A more general study \cite{DPVlong} of the run-away of the
dilaton during inflation (including an estimate of the 
effect of quantum fluctuations) only modifies this result by a factor ${\cal O}(1)$. 
It is also found that the present value of the dilaton is well approximated by 
$\varphi_{\rm end}$.

Eq.~(\ref{eq2.23'}) tells us that, within our scenario, the smallness of the present
matter couplings of the dilaton is quantitatively linked to the smallness of the
(horizon-scale) cosmological density fluctuations. To be more precise, and to 
study the compatibility with present experimental data, we need to estimate the
crucial dimensionless quantity
\be \label{eq3.1}
\alpha_A (\varphi) \equiv \partial \ln m_A (\varphi) / \partial \, \varphi \, ,
\ee
which measures the coupling of $\varphi$ to a massive particle of type $A$. 
The definition of $\alpha_A$ is such that, at the Newtonian approximation, the interaction 
potential between particle $A$ and particle $B$ is $-G_{AB} \, m_A \, m_B / r_{AB}$ where 
\cite{DN93,DP94} 
$G_{AB} = G (1 + \alpha_A \, \alpha_B) $.
Here, $G$ is the bare gravitational coupling constant entering the Einstein-frame action 
(\ref{eq2.4}), and the term 
$\alpha_A \, \alpha_B$ comes from the additional attractive effect of dilaton exchange.
 
Let us first consider the (approximately) composition-independent deviations from general 
relativity, i.e. those that do not essentially depend on violations of the equivalence 
principle. Most composition-independent gravitational experiments (in the solar system or in 
binary pulsars) consider the long-range interaction between objects whose masses are 
essentially baryonic (the Sun, planets, neutron stars). As argued in \cite{TV88,DP94} the 
relevant coupling coefficient $\alpha_A$ is then approximately universal and given by the 
logarithmic derivative of the QCD confinement scale $\Lambda_{\rm QCD} (\varphi)$, because the 
mass of hadrons is essentially given by a pure number times $\Lambda_{\rm QCD} (\varphi)$. [We 
shall consider below the small, non-universal, corrections to $m_A (\varphi)$ and 
$\alpha_A (\varphi)$ 
linked to QED effects and quark masses.] Remembering from Eq.~(\ref{eq1.1}) the 
fact that, in the string frame (where there is a fixed cut-off linked to the string mass 
$\widetilde{M}_s \sim (\alpha')^{-1/2}$) the gauge coupling is dilaton-dependent ($g_F^{-2} = 
B_F (\varphi)$), we see that (after conformal transformation) the Einstein-frame confinement 
scale has a dilaton-dependence of the form
\be
\label{eq3.5}
\Lambda_{\rm QCD} (\varphi) \sim C_g^{1/2} \, B_g^{-1/2} (\varphi) \exp [- 8 \pi^2 \, b_3^{-1} 
\, B_F (\varphi)] \, \widetilde{M}_s \, ,
\ee
where $b_3$ denotes the one-loop (rational) coefficient entering the Renormalization Group 
running of $g_F$. Here $B_F (\varphi)$ denotes the coupling to the ${\rm SU} (3)$ gauge 
fields. For simplicity, we shall assume that (modulo rational coefficients) all gauge fields 
couple (near the string cut off) to the same $B_F (\varphi)$. This yields the following 
approximately universal dilaton coupling to hadronic matter
\be
\label{eq3.6}
\alpha_{\rm had} (\varphi) \simeq \left( \ln \left( \frac{\widetilde{M}_s}{\Lambda_{\rm QCD}} 
\right) + \frac{1}{2} \right) \frac{\partial \ln B_F^{-1} (\varphi)}{\partial \, \varphi} \, .
\ee
Numerically, the coefficient in front of the R.H.S. of (\ref{eq3.6}) is of order 40. 
Consistently with our basic assumption (\ref{eq1.3}), we parametrize 
the $\varphi$ dependence of the gauge coupling $g_F^2 = B_F^{-1}$ as
\be
\label{eq3.7}
B_F^{-1} (\varphi) = B_F^{-1} (+ \infty) \, [1 - b_F \, e^{-c\varphi}] \, .
\ee
We finally obtain 
\be
\label{eq3.8}
\alpha_{\rm had} (\varphi) \simeq 40 \, b_F \, c \, e^{-c\varphi} \, .
\ee
Inserting the estimate (\ref{eq2.23'}) of the value of $\varphi$ reached because of the 
cosmological evolution, we get the estimate
\be
\label{eq3.9}
\alpha_{\rm had} (\varphi_{\rm end}) \simeq 3.2 \, \frac{b_F}{b_{\lambda} \, c} \, 
\delta_H^{\frac{4}{n+2}} \, .
\ee
It is plausible to expect that the quantity $c$ (which is a ratio) and the 
ratio $b_F / b_{\lambda}$ are both of order unity. This then leads to the numerical estimate 
$\alpha_{\rm had}^2 \sim 10 \, \delta_H^{\frac{8}{n+2}}$, with $\delta_H \simeq 5 \times 
10^{-5}$. An interesting aspect of this result is that the expected present value of 
$\alpha_{\rm had}^2$ depends rather strongly on the value of the exponent $n$ (which entered 
the inflaton potential $V(\chi) \propto \chi^n$). In the case $n=2$ (i.e. $V(\chi) = 
\frac{1}{2} \, m_{\chi}^2 \, \chi^2$) we have $\alpha_{\rm had}^2 \sim 2.5 \times 10^{-8}$, 
while if $n=4$ ($V (\chi) = \frac{1}{4} \, \lambda \, \chi^4$) we have $\alpha_{\rm had}^2 \sim 
1.8 \times 10^{-5}$. Both estimates are compatible with present (composition-independent) 
experimental limits on deviations from Einstein's theory (in the solar system, and in
binary pulsars).
For instance, the ``Eddington'' parameter 
$\gamma - 1 \simeq -2 \, \alpha_{\rm had}^2$ is compatible with the present best limits
$\vert \gamma - 1 \vert \lesssim 2 \times 10^{-4}$ coming 
from Very Long Baseline Interferometry measurements of the deflection of radio waves 
by the Sun \cite{exp}.

Let us consider situations where the non-universal couplings of the dilaton induce 
(apparent) violations of the equivalence principle. This means considering the 
composition-dependence of the dilaton coupling $\alpha_A$, Eq.~(\ref{eq3.1}), i.e. the 
dependence of $\alpha_A$ on the type of matter we consider.
Two test masses, made respectively of $A$- and $B$-type particles will fall in the 
gravitational field generated by an external mass $m_E$ with accelerations differing by
\be
\label{eq3.14}
\left( \frac{\Delta a}{a} \right)_{AB} \equiv 2 \, \frac{a_A - a_B}{a_A + a_B} \simeq 
(\alpha_A - \alpha_B) \, \alpha_E \, .
\ee
We have seen above that in lowest approximation $\alpha_A \simeq \alpha_{\rm had}$ does not 
depend on the composition of $A$. We need, however, now to retain the small 
composition-dependent effects to $\alpha_A$ linked to the $\varphi$-dependence of QED and 
quark contributions to $m_A$. This has been investigated in \cite{DP94} with the result
that $\alpha_A - \alpha_{\rm had}$ depends linearly on the baryon number $B \equiv N + Z$,
the neutron excess $D \equiv N-Z$, and the quantity 
$E \equiv Z (Z-1) / (N+Z)^{1/3}$ linked to nuclear Coulomb effects. Under the plausible 
assumption that the latter dependence is dominant, and using the average estimate 
$\Delta (E/M) \simeq 2.6$ (applicable to mass pairs such as (Beryllium, Copper)
or (Platinum, Titanium)), one finds that the 
 violation of the universality of free fall is approximately given by
\be
\label{eq3.17}
\left( \frac{\Delta a}{a} \right) \simeq 5.2 \times 10^{-5} \, \alpha_{\rm had}^2 \simeq 5.2 
\times 10^{-4} \left( \frac{b_F}{b_{\lambda} \, c} \right)^2 \, \delta_H^{\frac{8}{n+2}} 
\, .
\ee
This result is one of the main predictions of our model. If we insert the observed
density fluctuation 
 $\delta_H \sim 5 \times 10^{-5}$, we obtain a level 
of violation of the universality of free fall (UFF) 
due to a run-away dilaton which is
$\Delta a/a \simeq 1.3 ( b_F/(b_{\lambda} \, c) )^2 \times 10^{-12}$ for  $n=2$ (i.e.
for the simplest chaotic inflationary potential  
$V (\chi) = \frac{1}{2} \, m_{\chi}^2 (\phi) \, \chi^2$), and
$\Delta a/a \simeq 0.98 ( b_F/(b_{\lambda} \, c) )^2 \times 10^{-9}$ for $n=4$
(i.e. for $V (\chi) = \frac{1}{4} \, \lambda (\phi) \, \chi^4$).
The former case is naturally compatible with current tests 
(at the $ \sim 10^{-12}$ level \cite{Su94})
of the UFF. Values $ n \geq 4$ of the exponent require 
(within our scenario) that the (unknown) dimensionless combination of parameters 
$( b_F/(b_{\lambda} \, c) )^2$ be significantly smaller than one.

Let us also consider another possible deviation
from general relativity and the standard model: a possible variation 
of the coupling constants, most notably of the fine structure 
constant $e^2/\hbar c$ on which the strongest limits are available.
Consistently with our previous assumptions we expect 
$e^2 \propto B_F^{-1} (\varphi)$ so that, from (\ref{eq3.7}),
$e^2 (\varphi) = e^2 (+ \infty) \, [1 - b_F \, e^{- c \varphi} ]$.
The present logarithmic variation of $e^2$ (introducing the  derivative
$\varphi' = d \varphi / d p$ with respect to the ``e-fold'' 
parameter $dp = H \, dt = d a/ a $) is thus given by
\be
\label{eq3.21bis}
\frac{d \ln e^2}{H \, dt}  \simeq b_F \, c \, e^{-c \varphi} \, 
\varphi'_0 \, \simeq \frac{1}{40} \alpha_{\rm had} \varphi'_0 \, .
\ee
The current value of $\varphi'$,  $\varphi'_0$, depends on the coupling of the dilaton
to the two currently dominating energy forms in the universe: dark matter (coupling
$\alpha_m (\varphi)$), and vacuum energy (coupling $\alpha_V =  \frac{1}{4} \, \partial 
\ln V(\varphi) / \partial \, \varphi$). In the slow-roll approximation, one finds
\be \label{eq2.32bis}
(\Omega_m + 2 \Omega_V ) \varphi_0' 
= - \Omega_m \alpha_m - 4 \Omega_V \alpha_V.
\ee
where $\Omega_m $ and $\Omega_V$ are, respectively, 
the dark-matter- and the 
vacuum- fraction of critical energy density 
($\rho_c \equiv (3/2) {\widetilde{m}_P^2} H^2$). The precise value of $\varphi_0'$
is model-dependent and can vary (depending under the assumptions one makes) from
an exponentially small value ($ \varphi' \sim e^{- c \varphi}$) to a value of order
unity. In models where either the dilaton is more strongly coupled to dark matter
than to ordinary matter \cite{DGG}, or/and plays the role of quintessence (as suggested
in \cite{GPV}), $\varphi'_0$ can be of order unity. 
Assuming just spatial flatness and 
saturation of the ``energy budget" by
non-relatistic matter and dilatonic quintessence, 
one can relate the value of $\varphi' = d \varphi / (H dt)$ to 
$\Omega_m $ and to the deceleration parameter $q \equiv - \ddot{a}a/\dot{a}^2$:
$\varphi'^2 = 1 + q - 3 \Omega_m /2$. This yields the following generic, model-independent
relation between the present time variation of $e^2$, cosmological observables
and the level of UFF violation
\be
\label{eq3.26''}
\frac{d \ln e^2}{H \, dt} \simeq \pm \, 3.5 \times 10^{-6} \, 
\sqrt{  1 + q_0- 3\Omega_m/2  }
 \, \sqrt{ 10^{12} \frac{\Delta a}{a}} \, .
\ee
Note that the sign of the variation of $e^2$ is in general model-dependent ( as it
depends both on the sign of $b_F$ and the sign of $\varphi'_0$). Specific classes
of models might, however, favour particular signs of $d e^2/d t$. For instance,  
within the assumptions of \cite{V01} and \cite{GPV} it is natural to expect that
$e^2$ be currently {\it increasing}.

The phenomenologically interesting consequence of Eq.~(\ref{eq3.26''})
is to predict a time-variation of constants which may be large enough to be detected
by high-precision laboratory experiments. Indeed, using $H_0 \simeq 66$ km/s/Mpc, and the
plausible estimates
$\Omega_m = 0.3$, $q_0 = -0.4$,
Eq.~(\ref{eq3.26''}) yields the numerical estimate 
$d \ln e^2 / dt \sim \pm 0.9 \times 10^{-16}
\,   \sqrt{ 10^{12} \Delta a/a} \, {\rm yr}^{-1}$.
 Therefore, the current bound on UFF violations
($\Delta a/a \sim 10^{-12}$ \cite{Su94}) corresponds to the level $10^{-16}{\rm yr}^{-1}$,
which is comparable to the planned sensitivity of currently developed cold-atom clocks
\cite{salomon}. [Present laboratory bounds are at the  $10^{-14}{\rm yr}^{-1}$
level \cite{prestage,salomon}.] 
We note also that the upper limit on the variation of $e^2$ 
given by the Oklo data, i.e. 
$\vert d \ln e^2 / dt \vert \lesssim 5 \times 10^{-17} \, {\rm yr}^{-1}$ \cite{Oklo},
``corresponds''  to a violation of the UFF at the
level  $ \sim  10^{-13}$. Of course, present measurements are 
 not accurate enough to exclude an almost
 exact cancellation occurring in $ 1 + q_0- 3\Omega_m/2$: our work points at
 the relevance of establishing whether or not this can be excluded. 

We have also studied \cite{DPVlong} the variation of $e^2$ 
over cosmological times. By taking into account the constraints coming from the
need to be compatible with current cosmological data, we find that there is
no way, within our model, to explain a variation of $e^2$ as large as the 
recent claim \cite{Webb}
$ \Delta e^2/ e^2 = ( -0.72 \pm 0.18 ) \times 10^{-5}$ around redshifts $ z \approx 0.5-3.5$.
The largest possible variation we find (reached only if the UFF violation is just
below the $10^{-12}$ level, and if $\varphi$ is rather strongly coupled to dark matter),
is of order $ \Delta e^2/ e^2 = \pm 1.9 \times 10^{-6}$. This is only a factor $\sim 4$ below 
the claim \cite{Webb}
and is at the level of their one sigma error bar. Therefore a modest improvement in the
observational precision (accompanied by an improved control of systematics) will start to
probe a domain of variation of constants which, according to our scenario, corresponds
to an UFF violation smaller than the $10^{-12}$ level.

Our results suggest that the residual dilaton couplings today (as determined by a 
cosmological ``attraction'' towards the fixed point at infinite bare string coupling)
are just below the level $\alpha_{\rm had}^2 \sim 2.5 \times 10^{-8}$ corresponding
to a violation of the UFF at the 
$\Delta a / a \sim 10^{-12}$ level. This gives additional motivation for improved
tests of the UFF, such as the
Centre National d'Etudes Spatiales (CNES) mission MICROSCOPE \cite{Touboul} (to fly in 
2004; planned sensitivity: $\Delta a / a \sim 10^{-15}$), and  the National 
Aeronautics and Space Agency (NASA) and European Space Agency (ESA) mission STEP (Satellite 
Test of the Equivalence Principle; planned sensitivity: 
$\Delta a / a \sim 10^{-18}$)\cite{worden}. If our estimates are correct, these experiments
should find a rather strong violation signal.

 Another possible observable signal of a weakly coupled run-away dilaton is
 the time-variation of the natural constants. Here the conclusion depends crucially on the assumptions made about the
couplings of the dilaton to the cosmologically dominant forms of energy (dark matter and/or
dark energy). If these couplings are of order unity (and as large as phenomenologically 
acceptable),
the present time variation of the fine-structure constant is linked to the violation
of the UFF by the relation $d \ln e^2 / dt \sim \pm 2.0 \times 10^{-16}
\,   \sqrt{ 10^{12} \Delta a/a} {\rm yr}^{-1}$.
Such a time variation might be observable
( if $\Delta a/a$ is not very much below its present upper bound $\sim 10^{-12}$) through
the comparison of high-accuracy cold-atom clocks and/or via improved measurements of
astronomical spectra. The discovery of such a time variation (which is possible 
only if $(\Omega_m \alpha_m + 4 \Omega_V \alpha_V)/ (\Omega_m + 2 \Omega_V )$ 
is not too small or, in terms of more observable quantities,
 if $1 + q_0 - 3 \Omega_m / 2$ is of order 1), would then tell us that the dilaton plays an important cosmological role,
either because it is strongly coupled to dark matter 
($ \alpha_m \sim 1$) or/and 
because it plays the role of quintessence ($ \alpha_V \sim 1$).

G. V. wishes to acknowledge the support of a ``Chaire Internationale Blaise Pascal'', 
administered by the ``Fondation de L'Ecole Normale Sup\'erieure''.

\vspace{-2mm}

\begin{references}
\vspace{-6mm}
\bibitem{Witten}  E. Witten, {\it Phys. Lett. B} {\bf 149}, 351 (1984).
\bibitem{TV88} T.R. Taylor and G. Veneziano, {\it Phys. Lett. B} {\bf 213}, 450 (1988).
\bibitem{DP94} T. Damour and A.M. Polyakov, {\it Nucl. Phys. B} {\bf 423}, 532 (1994); {\it 
Gen. Rel. Grav.} {\bf 26}, 1171 (1994).
\bibitem{DN93} T. Damour and K. Nordtvedt, {\it Phys. Rev. Lett.} {\bf 70}, 2217 (1993); {\it 
Phys. Rev. D} {\bf 48}, 3436 (1993).
\bibitem{V01} G. Veneziano, {\it J. High Energy Phys.} {\bf 06}, 051 (2002)  
\bibitem{GPV} M. Gasperini, F. Piazza and G. Veneziano, {\it Phys. Rev. D} {\bf 65}, 023508
 (2002).
\bibitem{DV96} 
 T. Damour and A. Vilenkin, {\it Phys. Rev. D} {\bf 53}, 2981 (1996).
\bibitem{DPVlong} T. Damour, F. Piazza and G. Veneziano, hep-th/0205111.
\bibitem{L90} A. Linde, {\it Particle Physics and Inflationary Cosmology}, 
(Harwood, Chur, 1990). 
\bibitem{exp} D.E. Groom et al., {\it European Physical Journal} {\bf C15}, 1 (2000),
 see Chapter~14 ``Experimental tests of gravitational theory'',
 available on http://pdg.lbl.gov/; 
and C.M.~Will, {\it Living Rev. Rel.} {\bf 4}, 4 (2001).
1990).
\bibitem{DGG} T. Damour, G.W. Gibbons and C. Gundlach, {\it Phys. Rev. Lett.} {\bf 64}, 123 
(1990).
\bibitem{Su94} Y. Su et al., {\it Phys. Rev. D} {\bf 50}, 3614 (1994).
\bibitem{salomon}C. Salomon et al., in {\it Cold atom clocks on Earth and in space,  
Proceedings of the 17$^{th}$ Int. Conf. on Atomic Physics}, E. Arimondo, M.
Inguscio, ed., (World Scientific, Singapore, 2001), p 23;
Y. Sortais et al., {\it Physica Scripta}, {\bf 95}, 50 (2001). 
\bibitem{prestage} J. D. Prestage, R. L. Tjoelker, and L. Maleki, {\it
Phys. Rev. Lett.}
{\bf 74} 3511 (1995).
\bibitem{Oklo} A.I. Shlyakhter, {\it Nature} {\bf 264}, 340 (1976); T. Damour and F. Dyson, 
{\it Nucl. Phys. B} {\bf 480}, 37 (1996). 
\bibitem{Webb} J.K. Webb et al., {\it Phys. Rev. Lett.} {\bf 87}, 091301 (2001).
\bibitem{Touboul} P. Touboul et al., {\it C.R. Acad. Sci. Paris} {\bf 2} (s\'erie IV) 1271 
(2001).
\bibitem{worden} P. W. Worden, in {\it Proc. 7th Marcel Grossmann Meeting
on General Relativity}, R. T. Jantzen and G. Mac Keiser eds. (World Scientific, Singapore,
1996), pp 1569-1573.
\end{references}

\end{document}